# de Broglie Wavelength and Frequency of Scattered Electrons in the Compton Effect

Vinay Venugopal* and Piyush S Bhagdikar

[1]Division of Physics, School of Advanced Sciences
VIT University, Chennai Campus, Vandalur Kelambakkam Road
Chennai 600127, India
*Corresponding author. Email:vinayvenugopal@vit.ac.in

**Abstract**
The undergraduate courses on modern physics generally consider the particle interpretation of Compton effect. Motivated by a student's solution in an undergraduate examination on modern physics here we consider the wave characteristics of recoiled electrons in the Compton effect. The de Broglie wavelength, wave and clock frequency of the scattered electrons are expressed in terms of the wavelength and the frequency of the incident and the scattered photons respectively using the familiar particle interpretation of the Compton effect, where initially the electron is at rest and its spin is ignored. Both non-relativistic and relativistic cases are considered. Numerical values of de Broglie wavelength, wave and clock frequency of the scattered electron are calculated for an incident photon energy that was used in the original experiment of Compton as a function of the scattering angle of the recoiled electron. Considering the relativistic effects which are however insignificant for the de Broglie wavelength of the recoiled electron under these conditions, the minimum value obtained is in the range of X-rays. The non-relativistic de Broglie wave frequency obtained by neglecting the rest mass of the electron leads to an underestimation of its value. The implications of de Broglie wavelength and clock frequency for Compton scattering experiments are briefly discussed and possible extensions of the obtained mathematical formulations are indicated. The results are useful for understanding the wave-particle duality of the recoiled electron in the context of the Compton effect.

## 1. Introduction

In several undergraduate texts on modern physics [1,2], the concept of the particle nature of wave (light) is introduced prior to that of wave nature of the particle (electron). The latter is demonstrated by the Davisson-Germer experiment. The former is illustrated using the Compton effect where a photon collides with a stationary electron which is treated as a particle using the laboratory frame of reference and its spin is ignored. Rarely the concept of de Broglie waves are discussed in the context of Compton scattering. Several attempts have been made to obtain the expression for Compton shift by considering the interaction of electromagnetic waves with electron [3-5]. Schrödinger [4] considered the interaction classical electromagnetic waves with the de Broglie waves of the electron. Pedagogical exposition of Schrödinger's treatment was considered by Strand [6] and an approach similar to that of Schrödinger has been used by Su [7]. Compton Shift has also been explained as a double Doppler shift considering the interaction of electromagnetic wave train with electron [8]. But generally these are not discussed in the undergraduate courses on modern physics. de Broglie when attempting to obtain the relativistic transformation of Planck-Einstein's equation [9] proposed three different frequencies to a particle with rest mass $m_e$ (i) frequency of the internal energy of the particle at rest $\nu_C = m_e c^2 / h$ (ii) frequency of the total energy of the particle as measured by a 'fixed' observer: $\nu_{dB} = \gamma m_e c^2 / h$, where $\gamma$ is the Lorentz factor (iii) 'internal periodic phenomenon' /clock frequency as measured by a 'fixed' observer: $\nu_{cl} = \nu_C/\gamma$. Recently there is a renewed interest [10-13] to understand the internal clock frequency of the electron. The experiments of Catillon *et al.* [11], aimed at detecting the clock frequency of the electron in terms of interactions electrons with atoms indicate that one needs to consider the internal motion of the centre of charge of the electron around its centre of mass [10] with a frequency twice that of $\nu_C$ corresponding to the frequency of internal motion of the electron (zitterbewegung proposed by Schrödinger) used by Dirac.

This paper was motivated by a student's (the second author) solution to the problem posed in an undergraduate modern physics examination at VIT University, Chennai Campus, Chennai (August, 2011) on calculating the de Broglie wavelength of recoiled electrons in the Compton effect. The de Broglie wavelength and the de Broglie wave and clock frequency of the scattered electrons are generally expressed in terms of relativistic velocity of the recoiled electrons. Here we obtain these parameters in terms of the wavelength and frequency of the incident and the scattered photons. The familiar expressions for the conservation of energy and momentum that describe the Compton scattering by considering the particle nature of photon and electron (initially at rest and ignoring its spin) are used. Both non-relativistic and relativistic cases are considered. Considering an incident photon energy that was used in the original experiment of Compton [13], numerical values of the de Broglie wavelength, wave and clock frequency are obtained. The implications of de Broglie wavelength and clock frequency for Compton scattering experiments are briefly discussed and possible extensions of the obtained mathematical formulations are indicated. Hence these results can be effective for understanding the wave-particle duality of the recoiled electrons in the context of the Compton effect.

## 2. Results and Discussion

We obtain the wave characteristics of the scattered electrons from the familiar Compton scattering mechanics where the photon and the electron are treated as particles in the frame work of special theory of relativity. The electron before the collision is considered to be free (valid when the binding energy of the electron is negligible compared to the energy of the incident photon) and at rest in the laboratory frame of reference, and its spin is ignored. Figure 1 shows the geometry of Compton scattering in terms of the angles of the scattered photon ($\phi$) and the electron ($\theta$) respectively with respect to the direction of the incident photon.

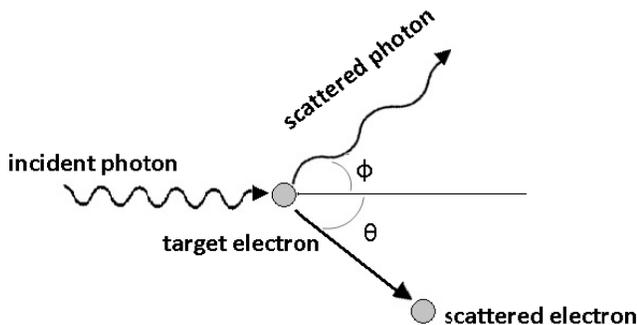

FIG. 1: The geometry of Compton scattering showing the directions of the scattered photon and electron with respect to the direction of the incident photon.

The relativistic de Broglie wavelength of the scattered electron can be obtained by applying the conservation of energy (Fig. 1),

$$p_{ph}c + m_e c^2 = \left(m_e^2 c^4 + p_e'^2 c^2\right)^{1/2} + p_{ph}' c \tag{1}$$

where $p_{ph}$ and $p_{ph}'$ are the momenta of the incident and the scattered photon respectively, $p_e'$ is the relativistic momentum of the scattered electron, $m_e$ is the rest mass of the electron and $c$ is the velocity of light. Substituting $p_{ph} = h/\lambda$, $p_{ph}' = h/\lambda'$ and $p_e' = h/\lambda_{dB}^R$ (where $h$ is the Planck's constant) respectively in Eq. (1),

$$\lambda_{dB}^R = \left(\left(\frac{1}{\lambda} - \frac{1}{\lambda'}\right)^2 + \frac{2}{\lambda_C}\left(\frac{1}{\lambda} - \frac{1}{\lambda'}\right)\right)^{-1/2} \tag{2}$$

where $\lambda_{dB}^R$ is the relativistic de Broglie wavelength of the scattered electron, $\lambda$ and $\lambda'$ are the wavelength of the incident and scattered photon respectively, $\lambda_C = h/m_e c$ is the Compton wavelength. A primitive form of this equation was derived by the student in the examination. A similar expression for the relativistic de Broglie wavelength of the scattered electron can be obtained using the conservation of momentum (Fig. 1)

$$\vec{p}_{ph} - \vec{p}_{ph}' = \vec{p}_e' \tag{3}$$

Taking the self vector dot product and substituting $p_{ph} = h/\lambda$, $p_{ph}' = h/\lambda'$ and $p_e' = h/\lambda_{dB}^R$,

$$\lambda_{dB}^R = \left(\frac{1}{\lambda^2} + \frac{1}{\lambda'^2} - \frac{2\cos\phi}{\lambda\lambda'}\right)^{-1/2} \tag{4}$$

where $\phi$ is the angle between the incident and the scattered photon. The de Broglie wavelength in Eq. (4) can be expressed in terms of the angle between the directions of the scattered electron with respect to the incident photon ($\theta$) using the well known relation,

$$\phi = 2\cot^{-1}\left[\left(1 + \frac{\nu}{\nu_C}\right)\tan\theta\right] \tag{5}$$

where $\nu_C = m_e c^2/h$.
Under the non-relativistic kinematics,

$$h(\nu - \nu') = \frac{1}{2}mV^2 \tag{6}$$

where $\nu$ and $\nu'$ are the frequencies of incident and scattered photons respectively, and $V$ is the velocity of the scattered electron. Hence the non-relativistic de Broglie wavelength of the scattered electron is

$$\lambda_{dB}^{NR} = \left(\frac{2}{\lambda_C}\left(\frac{1}{\lambda} - \frac{1}{\lambda'}\right)\right)^{-1/2} \tag{7}$$

The de Broglie wave frequency, associated with an electron is

$$\nu_{dB}^R = \frac{\gamma m_e c^2}{h} = \gamma \nu_C \tag{8}$$

where $\gamma$ is the Lorentz factor. Hence $v_C = m_e c^2/h$ with a value of $1.2356 \times 10^{20}$ Hz is the rest frequency that was identified by de Broglie with the internal clock of the electron. A non-relativistic expression for the de Broglie wave frequency is obtained by neglecting the rest mass of the electron and treating the scattered electron as a free particle. Hence $mc^2$ in Eq. (8) is replaced by the kinetic energy $E_k = p^2/2m$, where $p$ is the non-relativistic momentum of the scattered electron. Thus Eq. (8) becomes

$$v_{dB}^{NR} = \frac{p^2}{2mh} \tag{9}$$

Using $p = h/\lambda_{dB}^{NR}$ and substituting for $\lambda_{dB}^{NR}$ from Eq. (7) in Eq. (9),

$$v_{dB}^{NR} = v - v' \tag{10}$$

The expression for the relativistic de Broglie wave frequency in terms of the frequency of the incident and scattered photon is obtained starting from

$$v_{dB} = \frac{V_p}{\lambda_{dB}} \tag{11}$$

where $V_p$ is the phase velocity of the de Broglie waves associated with the scattered electron. Substituting for $\lambda_{dB}$ from Eq. (2), using $V_p = c^2/V$ and expressing $V$ in terms of $v$ and $v'$ using the relativistic kinetic energy of the scattered electron ($K = h(v - v')$), Eq. (11) becomes

$$v_{dB}^R = c \left(1 - \left(\frac{1}{\left(1 + \frac{(v-v')}{v_C}\right)^2}\right)\right)^{-1/2} \left(\frac{1}{c^2}(v-v')^2 + \frac{2m_e}{h}(v-v')\right)^{1/2} \tag{12}$$

Similarly using Eq. (4),

$$v_{dB}^R = \left(1 - \left(\frac{1}{\left(1 + \frac{(v-v')}{v_C}\right)^2}\right)\right)^{-1/2} \left(v^2 + v'^2 - 2vv'\cos\phi\right)^{1/2} \tag{13}$$

The internal clock frequency of the scattered electron as measured by an observer in the laboratory frame is

$$v_{cl} = \frac{v_{dB}^R}{\gamma^2} = \frac{v_{dB}^R}{\left(1 + \frac{(v-v')}{v_C}\right)^2} \tag{14}$$

where one should substitute for $v_{dB}^R$ using Eq. (12) or (13). Note that $\lambda'$ in Eq. (2), (4) and (7) can be expressed in terms of $\lambda$ and $\phi$ (or $\theta$) while $v'$ in Eq. (12), (13) and (14) can be expressed in terms of $v$ and $\phi$ (or $\theta$). Table I shows the calculated de Broglie wavelength (from Eq. (7) and (4)), de Broglie wave frequency (from Eq. (10) and (12)), and de Broglie clock frequency (from Eq. (14)) for an incident photon of wavelength 0.707831 Å using MATLAB R2008b. This wavelength was chosen for the purpose of illustration, since Compton in his original experiment [14] used MoK$\alpha_1$ X-ray source. For comparison, the value of $\nu_C$ is shown in the last column. The variation of $\lambda_{dB}^R$, $\nu_{dB}^R$, and $\nu_{cl}$ as a function of $\theta$ for an incident photon of wavelength $\lambda = 0.707831$ Å are shown in Fig. 2 obtained from Eq. (4), (12) and (14) respectively plotted using MATLAB R2008b.

TABLE I. The calculated de Broglie wavelength, wave frequency and clock frequency of the scattered electron for an incident photon of wavelength $\lambda = 0.707831$ Å at different angles of the scattered photon ($\phi = 45°$, 90°, 135°, 180°). The values of the constants $c$, $h$, $m_e$, $\lambda_C$, and the wavelength of MoK$\alpha_1$ radiation were obtained from the NIST website.[15]

| $\theta$ (°) | $\lambda_{dB}^{NR}$ (Å)[a] | $\lambda_{dB}^{R}$ (Å)[b] | $\nu_{dB}^{NR}$ (×10$^{17}$ Hz)[c] | $\nu_{dB}^{R}$ (×10$^{20}$ Hz)[d] | $\nu_{cl}$ (×10$^{20}$ Hz)[e] | $\nu_C$ (×10$^{20}$ Hz)[f] |
|---|---|---|---|---|---|---|
| 0 | 0.3658 | 0.3656 | 2.7173 | 1.2383073 | 1.2328786 | 1.2355899 |
| 21.82 | 0.3941 | 0.3939 | 2.3414 | 1.2379313 | 1.2332530 | 1.2355899 |
| 44.03 | 0.5090 | 0.5089 | 1.4037 | 1.2369936 | 1.2341878 | 1.2355899 |
| 66.81 | 0.9295 | 0.9294 | 0.4210 | 1.2360110 | 1.2351692 | 1.2355899 |

[a]Using (7), [b]Using (4), [c]Using (10), [d]Using (12), [e]Using (14), [f]$m_0c^2/h$

The calculated de Broglie wavelength of the recoiled electron in columns 2 and 3 of Table I, increases with the increasing angle of the scattered electron since the velocity of the recoiled electron is maximum at $\theta = 0$ and minimum at $\theta = \pi/2$. The value of the de Broglie wavelength in column 3 (also see Fig. 2 (a)) are lower than those of column 2 for a given $\theta$ indicative of relativistic effects. The asymptotic behavior of de Broglie wavelength (Fig. 2 (a)) close to $\theta = \pi/2$ is the result of assuming free electron at rest before collision. It is clear that under the conditions considered in Table I the relativistic effect on the de Broglie wavelength of the recoiled electron is not very significant since the maximum kinetic energy of the scattered electron under the assumed conditions (1.1238 keV) is small compared to its rest mass energy (511 keV). However at higher energies such as 0.1-1 MeV of incident photon, the relativistic effects cannot be ignored. The minimum de Broglie wavelength obtained in Table I (365.6 pm) is comparable to the wavelength of X-rays. Hence when the scattering medium is a crystalline solid the recoiled electrons under suitable conditions can undergo diffraction. de Broglie wavelength of recoiled electron has significance in Compton scattering experiments. A Compton resonance process has been reported where the recoiled electrons in Compton scattering resonate with the Si (111) crystal [16]. The de Broglie wavelength in a direction normal to the lattice plane is considered [16] for obtaining such resonance condition.

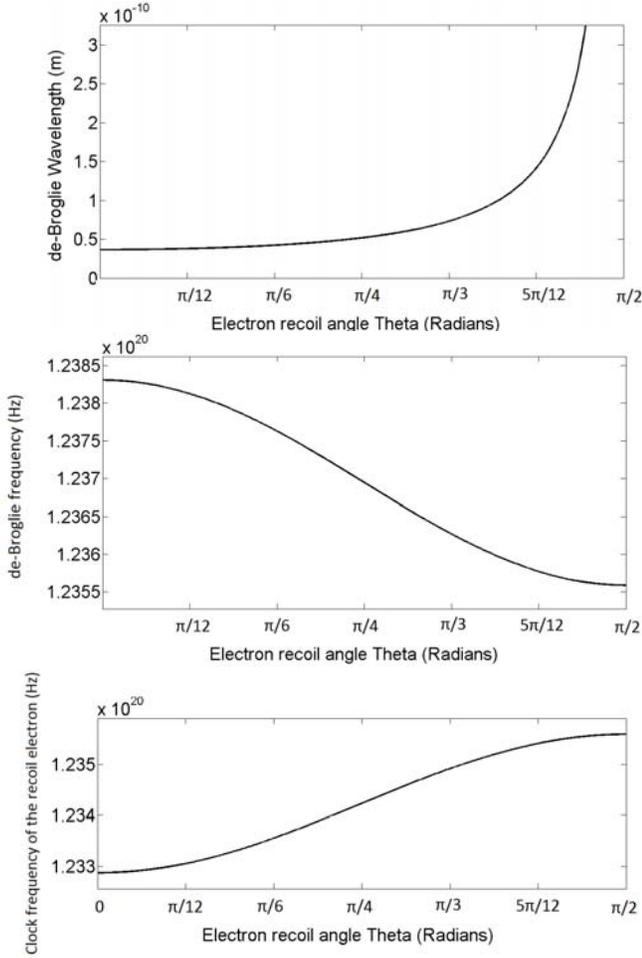

FIG. 2. The variation of (a) de Broglie wavelength (b) de Broglie wave frequency (c) de Broglie clock frequency of the scattered electron as a function of (a) scattering angle of the recoiled electron ($\theta$) using Eq. (4), (12), and (14) respectively for an incident photon of wavelength $\lambda$ = 0.707831 Å.

The de Broglie wave frequency of the scattered electron obtained by neglecting the rest mass of the electron (Eq. (10)) although consistent with literature [17] leads to an underestimation of the de Broglie wave frequency (see column 4 in Table I), since $\nu_{dB} = \nu_C$. For the relativistic case (column 5, Table I), the de Broglie wave frequency of the scattered electron decreases with increase of $\theta$ (see Fig. 2 (b)). However the increase is not very significant under the input parameters considered here ($(V/c)_{max} = 0.0662$). The de Broglie's clock frequencies ($\nu_{cl}$) of the scattered electron obtained from Eq. (14) are shown in column 6 of Table I. The clock frequency increases with increase of $\theta$ (see Fig. 2 (c)) again indicative of relativistic effects. The measurement of the internal clock frequency of the electron ($\nu_C$) using channeling of 80 MeV electron beam in silicon crystal has been reported [11]. The maximum kinetic energy of the scattered electron (1.1238 keV) under the conditions considered in Table I is highly insufficient for such channeling experiments. However the differential Klien-Nishina cross section $d_e\sigma/d\Omega_\theta$ as a function of $\theta$ indicates that at high incident photon energies (> 10 MeV) the recoiled electrons are very strongly forwarded directed along $\theta = 0$ and hence the possibilities of channeling of recoiled electrons at high incident photon energies (≥ 80 MeV) in crystalline scattering medium can to be explored.

The formulations mentioned in Eq. (2) and (14) can be useful to determine the de Broglie wavelength and frequency of the recoiled electrons while considering a guided photon [18] and when considering the effect of refractive index of the scattering medium on Compton scattering [19]. In addition these formulations can be extended to bound electron and when including the spin of the electron.

Teachers while evaluating students in the university examination must be receptive to new ideas or interpretations that can arise from student's answers/solutions. In an examination situation the student is forced to give an answer in a limited time. According to the experience of the first author of this paper who is involved in teaching physics and evaluating students, creativity of a student may be unknowingly triggered by the pressure to score good grades.

## 3. Conclusions

The de Broglie wavelength, wave and clock frequency of the scattered electron in the Compton effect were obtained in terms of the wavelength and frequency of the incident and the scattered photon respectively. The numerical values of these parameters for wavelength of incident photon that was used in the original experiment of Compton as a function of the scattering angle of the recoiled electrons indicate that (i) the relativistic effects for de Broglie wavelength is insignificant (ii) the minimum value of the de Broglie wavelength of the scattered electron calculated is in the range of wavelength of X-rays and hence under suitable conditions these recoiled electrons can undergo diffraction with the crystalline scattering medium (iii) the de Broglie wave frequency obtained by neglecting the rest mass of the electron leads to underestimation of its value. The relevance of the de Broglie wavelength and clock frequency of the recoiled electrons in Compton scattering experiments and future extensions of the obtained mathematical formulations were briefly highlighted.

**Acknowlegements**

The author (Dr.Vinay Venugopal) thanks VIT University, Chennai Campus, Chennai for providing a stimulating environment for innovative teaching and research infrastructure, and Dr. A. A. Samuel, Pro-Vice Chancellor for his encouragement. This work was presented as an oral contribution at the World Conference on Physics Education (WCPE)-2012, Istanbul, Turkey made possible with the financial assistance from VIT University, Indian National Science Academy (INSA), New Delhi, and IUPAP travel grant.

**References :**

[1] A. Beiser, S. Mahajan, and S. R. Choudhury, *Concepts of Modern Physics* (Sixth Edition, Special Indian Edition, Delhi: Tata McGraw-Hill 2009).
[2] P. A. Tipler & R. A. Llewellyn, *Modern Physics* (Sixth Edition San Francisco: W. H. Freeman & Co. 2012).
[3] C. Eckart, Phys. Rev. **24**, 591 (1924).
[4] E. Schrödinger, Ann. Phys. **28,** 257-264 (1927).
[5] C. V. Raman, Indian J. Phys. **3**, 357 (1928).
[6] J. Strnad, Eur. J. Phys. **7**, 217 (1986).
[7] C. C. Su, 'A wave interpretation of the Compton effect as a further demonstration of the postulates of de Broglie' Preprint physics/050621v3 (2006).
[8] See for example, R. Kidd, J. Ardini & A. Anton, Am. J. Phys., **53**, 641 (1985).
[9] L. de Broglie, C.R. Acad. Sci, **177**, 507 (1923).
[10] M. Rivas, 'Measuring the internal clock of the electron', Preprint physics/0809.3635v2 (2012).
[11] P. Catillon, N. Cue, M. J. Gaillard, R.Genre, M. Gouanère, R. G. Kirsch, J. –C. Poizat, J. Remillieux, L. Roussel, M. Spighel, Found. Phys. **38**, 659-664 (2008).
[12] D. Hestenes, Found. Phys. **40**, 1-54 (2010).
[13] G. R. Osche, Annales de la Fondation Louis de Broglie, **36**, 61-71 (2011).
[14] A. H. Compton, Phys. Rev. **21**, 483-502 (1923).
[15] See http://physics.nist.gov/cuu/Constants/
[16] K. D. Gupta, Phys. Rev. Lett., **13**, 338-340 (1964).
[17] R. Heyrovská, 'Compton shift and de Broglie frequency', Preprint physics/040104 (2004).
[18] G. R. Osche, Physics Essays, **21**, 260-265 (2008).
[19] S. G. Chefranov, 'New quantum theory of the Vavilov- Cherenkov radiation and its analogues', Preprint physics/1205.3774 (2012).
8